\documentclass[12pt]{article}

\usepackage{graphics,epsfig}

\def\dalemb#1#2{{\vbox{\hrule height .#2pt
        \hbox{\vrule width.#2pt height#1pt \kern#1pt
                \vrule width.#2pt}
        \hrule height.#2pt}}}

\let\a=\alpha  \let\g=\gamma  \let\e=\epsilon
  \let\th=\theta  \let\k=\kappa
\let\l=\lambda   \let\x=\xi  
\let\s=\sigma \let\t=\tau    
 
       \let\D=\Delta \let\Th=\Theta 
\let\X=\Xi  \let\S=\Sigma  \let\Y=\Psi
 
\let\la=\label \let\ci=\cite 
  
\def\nn{\nonumber} \def\bd{\begin{document}} \def\ed{\end{document}}
\def\ds{\documentstyle} \let\fr=\frac \let\bl=\bigl \let\br=\bigr
\let\Br=\Bigr \let\Bl=\Bigl
\let\bm=\bibitem
\let\na=\nabla
\def\tU{{\widetilde U}}
\let\pa=\partial \let\ov=\overline
\def\ie{{\it i.e.\ }}
\newcommand{\be}{\begin{equation}}
\newcommand{\ee}{\end{equation}}
\def\ba{\begin{array}}
\def\ea{\end{array}}
\def\ft#1#2{{\textstyle{{\scriptstyle #1}\over {\scriptstyle #2}}}}
\def\fft#1#2{{#1 \over #2}}
\def\F#1#2{{ F_{#1}^{(#2)} }}
\def\cF#1#2{{ {\cal F}_{#1}^{(#2)} }}

\def\R{{\bf R}}
\def\sst#1{{\scriptscriptstyle #1}}
\def\oneone{\rlap 1\mkern4mu{\rm l}}
\def\e7{E_{7(+7)}}
\def\td{\tilde}
\def\wtd{\widetilde}
\def\im{{\rm i}}
\def\bog{Bogomol'nyi\ }
\newcommand{\ho}[1]{$\, ^{#1}$}
\newcommand{\hoch}[1]{$\, ^{#1}$}
\newcommand{\bea}{\begin{eqnarray}}
\newcommand{\eea}{\end{eqnarray}}
\newcommand{\ra}{\rightarrow}
\newcommand{\lra}{\longrightarrow}
\newcommand{\Lra}{\Leftrightarrow}
\newcommand{\ap}{\alpha^\prime}
\newcommand{\bp}{\tilde \beta^\prime}
\newcommand{\cB}{{\cal B}}
\newcommand{\cO}{{\cal O}}
\def \ep {\epsilon}

\newcommand{\tr}{{\rm tr} }
\newcommand{\Tr}{{\rm Tr} }
\newcommand{\NP}{Nucl. Phys. }
\def\ve{\varepsilon}
\def\vf{\varphi}
\def\F{\Phi}
\def\wg{\wedge}
\def\thb{\bar{\theta}}
\def\Thb{\bar{\Theta}}
\def\barp{\bar{p}}
\def\barq{\bar{q}}
\def\barc{\bar{c}}
\def\bard{\bar{d}}
\def\e{\epsilon}

\def \bi{\bibitem}
\def \la {\label}

\def \l {\lambda}
\def\foot{\footnote}
\def \tl  {{\tilde \l}}
\def \sql {{\sqrt \l}}
\def \adss {$AdS_5 \times S^5$\ }
\newcommand{\rf}[1]{(\ref{#1})}
\def \ov {\over}

\def\th{\theta}
\def\Th{\Theta}
\def\vth{\vartheta}
\def\btheta{{\bar\theta}}
\def\ttheta{{{\tilde\theta}}}
\def\bttheta{{{\bar\ttheta}}}
\def\vth{\vartheta}

\def\ra{\rightarrow}
\def\N{{\cal N}}
\def\F{{\cal F}}
\def\ni{\noindent}

\def \ha{{1\ov 2}}

\def\Y{{\rm Y}}
\def\X{{\rm X}}
\def\tY{\tilde{\rm Y}}
\def\tX{\tilde{\rm X}}
\def\dY{\dot{\rm Y}}
\def\dX{\dot{\rm X}}

\def \J {\mathcal{J}}
\def \del {\partial}

\def\dF{\dot{F}}
\def\dG{\dot{G}}
\def\df{\dot{f}}
\def \E {{\cal E}}
\def \S {{\cal S}}
\def \J {{\cal J}}

\def\ms{\mathcal{S}}
\def\mj{\mathcal{J}}
\def\soj{\fr{\ms}{\mj}}
\def \R {{\bf R}}
\def \om {\omega}
\def \bE {\bar E}
\def \x {{\cal X}}

\def\pint{{-\!\!\!\!\!\!\int}}

\makeatletter
\let\old@startsection=\@startsection
\renewcommand{\@startsection}[6]{\old@startsection{#1}{#2}{#3}{#4}{#5}{#6\mathversion{bold}}}
\makeatother

\newcommand{\nln}{\nonumber\\}
\newcommand{\nl}{\nonumber\\&\hspace{-4\arraycolsep}&\mathord{}}
\newcommand{\nlnum}{\\&\hspace{-4\arraycolsep}&\mathord{}}
\newcommand{\earel}[1]{\mathrel{}&\hspace{-2\arraycolsep}#1\hspace{-2\arraycolsep}&\mathrel{}}
\newcommand{\eq}{\earel{=}}

\makeatletter \@addtoreset{equation}{section} \makeatother
\renewcommand{\theequation}{\thesection.\arabic{equation}}

\begin{document}
\overfullrule=0pt
\parskip=2pt
\parindent=12pt
\headheight=0in \headsep=0in \topmargin=0in
\oddsidemargin=0in

\vspace{ -3cm}

\thispagestyle{empty}
\begin{flushright}
ITEP-TH-12/05\\
PUTP-2151\\
UUITP-02/05 \end{flushright}
\begin{center}
{\Large\bf Matching quantum strings to quantum spins:
one-loop vs.~finite-size  corrections\par}

 \vspace{1cm} {
N. Beisert$^{a}$,
 A.A. Tseytlin$^{b,}$\footnote{Also at Imperial College London
 and  Lebedev  Institute, Moscow
 }
and  K.  Zarembo$^{c,}$\footnote{Also at ITEP, Moscow, 117259
Bol. Cheremushkinskaya 25, Russia} }\\
 \vspace{1cm}

{\it ${}^a$     Department of Physics, Princeton University,\\
Princeton, NJ 08544, USA
 }
 \vspace{.3cm}

{\it ${}^b$ Department of Physics, The Ohio State University,\\
Columbus, OH 43210, USA   }
 \vspace{.3cm}

{\it ${}^c$ Institutionen f\"or Teoretisk Fysik,
Uppsala Universitet\\
Box 803, SE-751 08 Uppsala,  Sweden}

\end{center}

\def \bi{\bibitem}
\def \la {\label}

\def \l {\lambda}
\def\foot{\footnote}
\def \tl  {{\tilde \l}}
\def \sql {{\sqrt \l}}
\def \adss {$AdS_5 \times S^5$\ }
\def \ov {\over}

\def \varpi {{\rm w}}
\def \S {{\cal S}}
\def \rk {k}

 \vspace{0.1cm}

 \begin{abstract}
 \noindent
We compare quantum corrections to semiclassical spinning strings in
$AdS_5 \times S^5$ to one-loop anomalous dimensions in ${\cal N} =4$
supersymmetric gauge theory. The latter are computed using the
reduced (Landau-Lifshitz) sigma model and with the help of the Bethe
ansatz. The results of all three approaches are in remarkable
agreement with each other. As a byproduct we establish the
relationship between linear instabilities in the Landau-Lifshitz
model and analyticity properties of the Bethe ansatz.
\end{abstract}
\newpage

\setcounter{equation}{0}
\setcounter{footnote}{0}
\setcounter{section}{0}

\def \tl {\tilde \l} \def \D {\Delta}

 \def \om {\omega}
\def \t {\tau}
\def \s {\sigma}
\def \X {{\rm X}}
\def \LL {Landau-Lifshitz\ }
\section{Introduction}

Recently,  there was a remarkable progress towards uncovering the structure of the  spectrum
of energies of (non-interacting)  quantum strings in $AdS_5 \times S^5$ or,
equivalently, the spectrum of dimensions of
single-trace operators in the dual $\N=4$ SYM theory
with $N\to \infty, \ \l= g^2_{\rm YM} N =$fixed
(for reviews see, e.g., \ci{beis,ts1,ts2,zare}).
Both energies $E$ and dimensions $\D$ depend on the `t Hooft coupling $\l$
(the string tension is $T = { \sql \ov 2 \pi}$) as well as on quantum numbers
like spins $J$ and ``winding{}'' numbers $m$  characterizing the states,
and one  basic implication of the AdS/CFT duality is the equality of the functions
$E(\l, J,m,...) = \D(\l, J, m,...)$  for {\it any}  value of the arguments.
It is not a priori clear how such a  relation can be tested for far-from-BPS states,
but remarkably  it was found that both the perturbative string theory and
the perturbative gauge theory  (where
anomalous dimensions are described by a spin
 chain)
contain certain states
for which the large $J$  expansions
of $E$ and $\D$ have similar
structures (see, e.g.,
 \ci{ft2,afrt,art} and
 \ci{bmsz,bfst,mez,ss,char,minah})\foot{Remarkably, the spectrum
 of the corresponding ferromagnetic spin chains does contain
``macroscopic string''
states \ci{old,bmsz}
 for which the spin chain
energies or anomalous dimensions
  scale as $\l^\ell \ov J^{2\ell-1}$  at
   $\ell=0,1,2,...$ loop orders.
}
\be \la{st}
E= J \bigg[1 + { \l \ov J^2} ( c^{(1)}_{0}  + { c^{(1)}_{1} \ov J} + { c^{(1)}_{2} \ov J^2} + ... )
+ { \l^2 \ov J^4} (  c^{(2)}_{0}  + { c^{(2)}_{1} \ov J} + { c^{(2)}_{2} \ov J^2} + ... ) + ...
\bigg]
\ , \ee
\be \la{ga}
\D= J\bigg[1  + { \l \ov J^2} ( a^{(0)}_{1}  + { a^{(1)}_{1} \ov J} + { a^{(2)}_{1} \ov J^2} + ... )
+ { \l^2 \ov J^4} (  a^{(0)}_{2}
 + { a^{(1)}_{2} \ov J} + { a^{(2)}_{2} \ov J^2} + ... ) + ...
\bigg]
\ .  \ee
The coefficients  $c^{(n)}_{\ell}$ (corresponding to $\ell$-loop
string-theory  correction $\sim ({ 1 \ov \sql})^{\ell+1}$,
\ $J= \sql \J$)
and $ a^{(n)}_{\ell}$   (corresponding to $\ell$-loop
gauge-theory  correction $\sim \l^\ell$)
 depend  on ratios
of spins and other quantum numbers and are finite in the large $J$ limit.
Similar expressions are found for near-BPS states
describing small string fluctuations near the
BMN vacuum state.\foot{
In this case \ci{par,callan}
one is to extract one  power of $J$, i.e. $
E= J  + { \l \ov J^2} ( c^{(1)}_{1}  + { c^{(1)}_{2} \ov J}
+ { c^{(1)}_{3} \ov J^2} + ... )
+ { \l^2 \ov J^4} (  c^{(2)}_{1}  + { c^{(2)}_{2} \ov J} +
{ c^{(2)}_{3} \ov J^2} + ... ) + ...$, etc.}

The two expressions, however, are obtained in the two different
limits.  On the
 string side one uses semiclassical expansion in which $\J = { J \ov \sql}$ is kept fixed
while  one first expands in $\a' \sim {1 \ov \sql}$ or, equivalently, in $ 1 \ov J$;
one may  then  also expand in $\tl = { 1 \ov \J^2}= { \l \ov J^2} $,
 which corresponds to  studying ``fast-moving{}''
strings. On gauge-theory side, one uses standard planar perturbation theory,
i.e.~first expands in $\l$ and then may also expand the resulting $l$-loop anomalous
dimensions in  large $J$ (or, equivalently, for the simplest cases
we are going to consider)  in large length of the operator.
What is even more surprising (given that the string and gauge theory  limits are ``opposite{}''
as far as $\l$ is concerned while  $E=\Delta$ should in general
 contain
non-trivial  interpolating functions of $\l$)
is that the first two  leading coefficient functions
\rf{st} and \rf{ga}  were found to be exactly the same:
\be \la{ma}
c^{(1)}_{0} = a^{(0)}_{1} \ , \ \ \ \ \ \ \
c^{(2)}_{0} = a^{(0)}_{2} \ , \ee
i.e.~the  two leading classical  string theory coefficients  match the
two leading (one- and two-loop) gauge theory coefficients.

This matching can be demonstrated in a very general way
by either extracting the  corresponding   coherent state (``Landau-Lifshitz{}'')
sigma model describing low-energy states of the ferromagnetic spin chain
and matching it to a ``fast-string{}'' limit of the
classical superstring action \ci{kru,krt,hl1,st,mik12,kt,mik3,hl2},
or by matching the integral  equation for
 the general solutions
of the integrable string  sigma model  to a similar equation for the density of the Bethe root
distribution appearing on the spin
chain side \ci{kmmz,kz,bksa,af,sak}.
This matching applies
  also  to near-by fluctuations, and is of a novel
truly ``microscopic{}''
or ``dynamical{}'' nature, i.e.~it appears to go beyond of a kind of matching based on
a non-renormalization
 theorem or BPS-saturation of a  coefficient of a certain term in a
 gauge or supergravity  effective action familiar in other (e.g, in matrix theory)
 contexts.

An explanation of why one gets this
  agreement for the two leading  coefficients
but apparently $c^{(n)}_{0} \not= a^{(0)}_{n}, \ n >2$
\ci{ss,bds}
(so that one needs to resum the series to verify that $E=\D$)
may  be traced to the structure of the dilatation operator
(or 1-d S-matrix \ci{s})
on the gauge theory side.
The structure of the dilatation operator (best
understood so far
in $SU(2)$ \ci{mz,bks,bds}  or $SU(2|3)$ \ci{b2}
sectors)
 is    dictated
 to a large extent by
the  maximal supersymmetry.
The   observed matchings
(for near-BMN  states to two orders
in $\l\ov J^2$ and first order in $1\ov J$;
for spinning string states to  two leading
orders in $\l\ov J^2$)
can be attributed to the fact that  the
the  ``gauge-theory{}'' \ci{bks}
 and the ``string-theory{}'' \ci{afs,b}
Bethe ansatze start to disagree
 at $\l^3$ order, while both
should be   limits of  a ``Better ansatz{}'' \ci{afs}
that contains  interpolating
 functions of $\l$ and $J$.


This  suggests that one should expect more
 matching between the coefficients
in \rf{st} and \rf{ga} at the {\it first two} orders in $\l$, in
particular, \be \la{mo} c^{(1)}_{1} = a^{(1)}_{1} \ ,  \ee i.e.~the
{\it 1-loop }  quantum string theory correction to the semiclassical
string energy  should match the leading finite-size correction to
the {\it 1-loop}
 gauge-theory anomalous dimension.
This is a novel situation since  (like
near-BMN example) it involves
 quantum string theory result  (incorporating, in particular,
  fermionic
  contributions) while previous
tests where for purely bosonic classical string
solutions.\foot{The possibility of matching of the leading
classical coefficients \rf{ma} does depend implicitly on the
full structure of the quantum superstring theory:
this matching depends on the fact that quantum string
 corrections are
suppressed in the large $J$ limit,
which  itself is a consequence of
the 2-d conformal invariance and underlying supersymmetry
of the superstring sigma model  \ci{ft2,ft3}.}

Attempts to test \rf{mo} were made  previously in \ci{fpt,ptt}
where  string 1-loop corrections to energies of particular circular
strings  were  computed and
 were  compared  to the spin
chain results for the $1/J$ corrections found  in \ci{lz,kz}.
 An apparent
disagreement was reported:
 it appeared that only the zero-mode string theory  result
for the 1-loop correction to the energy (given by a familiar sum of
the fluctuation mode frequencies) was captured by the Bethe ansatz
\ci{lz,kz}.
We shall compute an additional anomalous contribution to the Bethe ansatz%
\footnote{We would like to thank V.~Kazakov for inspiring discussions
on possible anomalies in the Bethe equations.}
overlooked in \ci{lz,kz}, see Appendix~\ref{appsadlkfj},
which restores the agreement with the string  calculation.

 In section 2 we shall review the 1-loop string results
 of \ci{ft3,fpt,ptt} for the leading  $1/J$ correction
 to the energy of circular spinning strings. We shall
  explain in particular that the full superstring 1-loop
  correction to the string energy
   can be understood
  as a $\zeta$-function regularized expression
  for the 1-loop correction to the string soliton energy
  computed  directly
  in the corresponding ``reduced{}'' or ``Landau-Lifshitz{}''
  sigma-model (i.e.~in the continuum limit of
  the coherent state path integral
  corresponding to the spin chain  Hamiltonian)\footnote{The
quantization of the classical solutions in the \LL model with the
help of zeta-regularization was considered before by J.~Minahan
(unpublished). We would like to thank him for the discussion of
these results.}.
  Then in section 3  we shall show how a  careful account of an
  anomaly (discussed in Appendix \ref{appsadlkfj})
  appearing in the finite-size expansion of the Bethe ansatz
  equations leads to the same  expression as
  found on the string side.

  Given that the string expression
   is essentially the
  regularized result of the  \LL model  and that
  the energies of
  the  small-fluctuation \LL modes  can be
  reproduced \ci{bmsz}
   on the Bethe ansatz side, this may seem to make
   the agreement quite natural (modulo the fact
    that the two
   computations  still apply in two different limits).
   This   stresses again a ``microscopic{}''
    nature of the matching:
not only the final expressions for the
coefficients match but also there is a remarkable
correspondence  between  the
intermediate steps in the respective calculations.
Surprisingly, the \LL  model  continues
to provide a conceptual link between  the string theory and
the spin chain even beyond the leading semiclassical
approximation.

Let us mention also that our result  may
shed more light on the
 AFS \ci{afs} ansatz  which is  {\it a} ``discretization{}''
 of the classical  string sigma model  solution.
As was shown  in \ci{b}, the AFS ansatz gives rise to
a spin chain at small coupling $\l$.
This spin chain agrees precisely, including all
$1/J$ effects, with gauge theory up to two loops.
The present work  gives support to the idea that
the quantum string at small $\l$ is indeed
described by a spin chain, be it the one
of gauge theory \ci{mz,bks}
or the string chain of \ci{b}
(which are both equivalent to the Heisenberg spin chain at this order).


\section{One-loop   superstring  vs.~quantum\\
 Landau-Lifshitz model}

\subsection{ $SU(2)$ case  }

It is best to start with the  simplest possible two-spin
solution in
the $SU(2)$ sector \ci{ft2}: rigid circular
 string rotating  with two
equal angular momenta $J_1=J_2 = \ha J $ in $S^3$ part of $S^5$.
In the form given  in \ci{art} (equivalent to the solution
of \ci{ft2}
by an $SO(4)$ rotation) it is $t = \k \t$, \
X$_1= \cos \psi\  e^{i \phi_1} = { 1 \ov \sqrt 2} e^{ i w \t + i k \s},$ \
 X$_2=\sin \psi\  e^{i \phi_2} = { 1 \ov \sqrt 2} e^{ i w \t - i k \s}.$ \
 Here $|\X_1|^2 +|\X_2|^2 =1$,
  $w = \J = { J \ov \sql}, \  \k^2 = \J^2 + k^2$ and
  $k$ is an integer  winding number. The classical energy $E = \sql \k$
  of this solution is \ci{ft2}  (cf.  \rf{st})
  \be
   E = \sqrt{ J^2 + \l k^2}
  =  J \big(1  + { \l k^2 \ov 2 J^2}  + ... \big) \ .
  \la{cla} \ee
To find the string 1-loop  correction to $E$ one is supposed
to determine  the characteristic frequencies $\om_n$
of the  bosonic and
fermionic fluctuations $\sim e^{i \om_n \t + i n \s}$
 and then compute an appropriate sum over them.
Considering the $AdS_5$ time $t$  and
the  ``fast{}'' motion direction
$\ha(\phi_1 + \phi_2)$ as the ``longitudinal{}'' directions, there will be
8 bosonic and 8 fermionic fluctuations. Among the bosonic ones, the
remaining  two of  $S^3$ fluctuations
 ($\psi$ and $\ha(\phi_1 - \phi_2)$)
 play a special role compared to four  $AdS_5$ and two  other $S^5$
fluctuations): they  belong to the $SU(2)$  \LL sigma-model
\ci{kru,krt}.
  Using the results of \ci{ft3}
(summarized in Appendix B of \ci{fpt}) we then find
 the
expression for the 1-loop  correction as a sum
of the zero ($n=0$ or constant in $\s$)  mode
and non-zero mode contributions\foot{We are using here  that in
the case of the ``homogeneous{}''
solutions of \ci{art}   the isometric angles are linear
functions of $\t$ and $\s$ and so  their
derivatives (and thus all coefficients in the fluctuation Lagrangian)
 are constant.
Also,
the connection in the fermionic covariant derivative is
automatically constant (one does not need a $\s$-dependent
rotation considered in \ci{ft3}), and thus  the fermions
are periodic and
their
modes are labelled by the integers $n$ just like
 as the bosonic modes.}
\be
  E_1 = E_{\rm  zero} + E_{ \rm non-zero} \ , \ \ \ \ \ \ \ \ \ \
  E_{1\ \rm non-zero}=  \sum^\infty_{n=1}  S_n \ , \la{su}
  \ee
  \be
E_{ \rm zero} = 2 +  \sqrt{1 -{ 2 \rk^2\ov \J^2 + k^2}  }
- 3  \sqrt{1 -{  \rk^2\ov \J^2 + k^2}  }
\ ,\la{dou}  \ee
\bea
 S_n
  =&& 2 \sqrt{ 1 +
   { ( n  + \sqrt{n^2 - 4 \rk^2   } )^2 \ov 4 ( \J^2 + k^2)}    }
+{2}\sqrt{1 + {n^2-2\rk^2\ov \J^2 + k^2 } }
 +{4}\sqrt{ 1 + {n^2\ov \J^2 + k^2 } } \nn\\
 &&
-{8}\ \sqrt{1   +  { n^2-\rk^2 \ov \J^2 + k^2 } }
 \ . \la{hi}
 \eea
The  sum  over $n$ is   finite as a consequence of the
conformal invariance  of the  string theory \ci{MT,ft2,ft3}:
the bosonic and fermionic divergences cancel each other.
The first (``fourth root{}'')
term in $S_n$ in \rf{hi} is the contribution of the two  ``transverse{}''
$S^3$ fluctuation modes. Note that the latter with  $n < 2 k$
   are tachyonic  \ci{ft2}
 and thus contribute
to an imaginary part of $E_1$. We shall ignore this
problem as a  very similar discussion will
apply also in the stable $SL(2)$ case
considered in \ci{ptt} and below; moreover, we will still  be able
to formally  match this string result to the
$SU(2)$ spin chain one despite this
instability problem.

Expanding the above expression at large $ \J$ or small
$\tl = { 1 \ov \J^2} =  { \l \ov J^2}$ we find:
 \be\la{z}
E_{ \rm zero}  =    {\l  k^2 \ov 2 J^2}   +  O (  { \l^2 \ov J^4 } )
\  , \ee
\be \la{ha}
E_{ \rm non-zero}
 =   { \l \ov 2 J^2 } \sum^\infty_{n=1}  \
 \big( n \sqrt{n^2 -4k^2 }  - n^2 +  2 k^2   \big)  +
  O (  { \l^2  \ov J^4 } )  \ . \ee
Note that  the leading term in  the zero-mode contribution \rf{z}
is exactly  the same as in the classical energy \rf{cla} but
with an extra $1 /J$ factor. This appears to be a universal feature:
we will find the same  in the $SU(3)$ and $SL(2)$ sectors.

A justification for using the $1/\J$ expansion before doing the infinite
sum is that the sum in \rf{ha} is again convergent.\foot{We have
checked  numerically (as in \ci{fpt})
that the finite sum in  \rf{ha}  matches indeed
the $1/\J^2$ coefficient in
the expansion of the function
obtained by computing  first the sum in \rf{su}
(for $k=1$  we got  $E_{\rm zero} +
E_{\rm non-zero} \approx \tl  ( - 0.4667 +  0.866 i )$
as in \ci{fpt}).
In general,
one may wonder why  one can obtain the $1/\J$ coefficient
 by first expanding in $1/\J$ and then doing the
sum over $n$ since $n$  can be bigger than $\J$. Indeed,
 the  coefficients  of the
higher-order $1/ \J^{4}$, etc.  terms in the expansion
 are given by
divergent series. What happens is  that a
resummation of the divergent part of the
$1/\J$ expansion that makes the result finite (as it was
originally in \rf{su} before expanding frequencies in $1/\J$)
 should not change the coefficient
of the leading finite $1/ \J^2$ term.
}
Notice that the first $n \sqrt{n^2 -4k^2 }$
 term in  \rf{ha} originated from the  large $\J$ expansion
 of the first  term in the sum in \rf{hi}
 (which is equal to the sum of the two $S^3$ fluctuation
 frequencies)
 and is thus  the contribution of the \LL fluctuation mode.
The other two terms $ -n^2$ and $ 2 k^2$ that make the sum
in \rf{ha} finite receive contribution from all the bosonic
{\it and} fermionic modes that thus
conspire to make the sum finite.

\bigskip

Let us now  compare this superstring  result  with what
one finds using an apparently  naive  procedure
based  quantization  of the \LL sigma model.
{}From the $SU(2)$ spin chain  perspective,
one  first  replaces the quantum mechanics of the Heisenberg
ferromagnet  by a coherent-state path integral
with the action containing the coherent-state
expectation value of the Hamiltonian, i.e.~$ <n| H | n> =
{\l\ov 16 \pi^2}
\sum^J_{a=1} ( \vec n_{a+1} - \vec n_a )^2 $,
where $   <n| \vec \s_a | n> = \vec n_a$.
As discussed in \ci{kru,krt}, concentrating on a particular class
of low-energy coherent states one can then define
a semiclassical limit of the coherent state  path integral
as $J \to \infty$ with $ \tl = {\l \ov J^2}$ fixed.
Indeed, in  this limit one is able to take the continuum limit
of the action,\foot{Note that it is necessary to  take
the continuum limit in order to define the semiclassical expansion:
only then the factor of  $J$ appears in front of the action and thus plays the role of the
inverse Planck's constant.}
   ending up
with the \LL action
for a unit 3-vector 2d field $\vec n(\tau,\s)$ \be\la{acc}
I=   J \int d\t \int^{2\pi}_0 {d \s\ov 2\pi}\
 \big[\ \vec C(n) \cdot \dot
{\vec n} \  - \ { 1 \ov 8}\  \tl\   \vec n' \cdot \vec n'\ \big]\ . \ee
This action happens to be the same \ci{kru}
as the ``fast-motion{}'' ($\tl \to 0$) limit of the relevant $R_t \times
S^3$ bosonic part of the \adss string action, demonstrating, in
particular,
the matching of the  leading-order coefficients in \rf{ma}.
 To go beyond the leading classical approximation would mean
 to include the leading $1/J$, i.e.~one-loop, correction
 to the coherent state  path integral.

On the spin chain side,  given that
we have first taken the continuum
 limit and {\it then}
 want to include quantum corrections,
  it is not a priori clear
 that  quantization of the  \LL model
  is going to   reproduce
 the finite-size or $1/J$
  expansion of the Bethe ansatz solution.
On the string side, first taking the
 ``fast string{}'' limit and then
quantizing the resulting reduced
 bosonic sigma model (thus ignoring quantum
 fluctuations in other directions)
  appears bound to be
wrong,  since, in particular, one needs
 the fermionic contributions
to make the 1-loop correction finite.\foot{Modulo the
UV divergence problem, one could try to justify this
using the effective field theory  philosophy
observing that other fluctuations should be
heavy in the large
$\J$ limit.}

And yet, this naive procedure manages  to reproduce
essentially
the right
answer for the leading $\l/J^2$ correction: it just needs
to  be
supplemented by a specific regularization prescription
dictated  by the underlying  microscopic theory
(spin chain described by Bethe ansatz  or quantum superstring).
 It turns out that the  Bethe
ansatz result dictates that this regularization
should be the standard $e^{-\ep n}$ or
$\zeta$-function regularization. Moreover,
this prescription happens to produce
 exactly the same result as the full superstring  calculation --
 the role of other superstring modes
  happens to reduce just  to making  the sum
 of the \LL frequencies finite!

\bigskip

To see  how this happens in some  detail,
 let us start with the \LL equation  on $R_\tau \times S^1_\s$
 which follows from \rf{acc}
\be\la{ll}
\dot   n^i = { 1 \ov 2 } \tl\  \epsilon^{ijk} n_j  n{}''_k \ ,
 \ \ \ \ \ \   \vec n^2=1 \ , \ \ \ \
 \tl \equiv  {\l \ov J^2}\ ,
\ee
and consider the simplest non-trivial
static solution  corresponding to the circular string
with $J_1=J_2= \ha J$
\be \la{ko}
 n_i = ( \cos 2k \s, \ \sin 2k \s,\  0) \ ,\ \ \ \ \ \  k=1,2,...
 \ .     \ee
 Expanding \rf{ll} near this solution with
 small perturbations parametrized by  two independent  functions
 $A_1,A_2$
\be
\delta n_i = ( - \sin  2k \s \ A_1(\tau,\s),
 \ \cos 2k \s\ A_1(\tau,\s), \ A_2(\tau,\s)) \ , \ \ \ \
 n_i \delta n_i=0 \ , \ee
 it is easy to show that
$ \dot A_1 = -\ha \tl ( A_2{}''  + 4 k^2 A_2)$,
$\dot A_2 = \ha \tl A_1{}''$.
Expanding the fluctuations in modes
  $A_s \sim \sum_{n =-\infty}^\infty
  C_{s,n}  e^{i w_n  \tau + i n \s} $ we  find that  the
characteristic frequencies  are  given by
\be \la{hc}
w_n = \pm \ha \tl     n \sqrt{ n^2 - 4 k^2} \ . \ee
These  fluctuation energies   were indeed
reproduced (for   $n > 1$) also  from the
  Bethe ansatz in \ci{bmsz}.

As in the standard  quantum oscillator case, the
correction to the classical energy (cf. \rf{cla};
superscript $^{(1)}$ indicates that
 this is order $\tl$ contribution)
\be \la{ene}
E^{(1)}_0 =   { 1 \ov 8}  \tl  J \int^{2\pi}_0
{ d\s\ov 2 \pi}    n'_i n'_i  =
\ha { \tl J  k^2}  = { \l k^2 \ov 2 J}  \ee
should then be  given by
\be\la{one}
E^{(1)}_1 = \ha \sum^{\infty}_{n=-\infty} |w_n|
=\ha \tl
 \sum^{\infty}_{n=1}   n \sqrt{ n^2 - 4 k^2} \ . \ee
 This sum is divergent, but let us compute it
   by first
  adding and subtracting  the divergent part
  and then renormalizing  the divergent part
   using the  $e^{-\epsilon n}$
 regularization and
 dropping terms singular in the  $\epsilon
 \to 0$ limit (this  is  equivalent
 to the $\zeta$-function regularization
 prescription). We get
 \be \la{fi}
 E^{(1)}_1 =   E_{\rm reg}   + E_{\rm fin}\ , \ \ \ \ \
\ \ \  E_{\rm reg} =
{ \l \ov 2 J^2}
 [\sum^{\infty}_{n=1}  ( n^2 - 2 k^2 )]_{\rm reg}  \ , \ee
\be \la{po}
 E_{\rm fin}= { \l \ov 2 J^2}
 \sum^{\infty}_{n=1}
 ( n \sqrt{ n^2 - 4 k^2}  -   n^2 + 2 k^2 ) \ . \ee
 The finite part of the
 regularization-dependent term  $E^{(1)}_{\rm reg}$   is
 then\foot{We use that
 $[\sum^\infty_{n=1}  n^2]_{\rm reg} =\zeta(-2) =0, \
 [\sum^\infty_{n=1}1]_{\rm reg} = \zeta(0)=-\ha$.
 Note also  that the divergence coming
 from the $n^2$  term could be ruled out by the condition that the
 result should vanish in the case of $k=0$  which corresponds to
 fluctuations near the BMN vacuum $\vec n = ( 1,0,0)$.}
 \be \la{jo}
 E_{\rm reg} =  \ha \tl k^2 = { \l k^2 \ov 2 J^2} \ .
 \ee
 Comparing \rf{jo}  and \rf{po}  with the string results
 \rf{z} and \rf{ha} we conclude that

 (i)  the regularized  value
 $E_{\rm reg}$ of the divergent part of the sum
  of the \LL fluctuation modes happens to be
  the same as the leading
 zero-mode contribution, and

(ii) the finite sum of {\it all} non-zero mode
 string contributions
 turns out  to be equal just to the finite part
$E_{\rm fin}$ of the sum of the  \LL modes.

The  string theory thus provides
an automatic regularization of the \LL mode
contributions. Given that the \LL fluctuations  are ``visible{}''
\ci{bmsz,frey} on the spin chain side
this gives a strong hint that there should be
a precise matching between the  quantum string
and the  spin chain results for the $1/ J$ correction.
 As we shall explain in section 3 below,
the one-loop anomalous dimension computed from the Bethe ansatz
equations indeed agrees
 with the string calculation.


\bigskip

The above expressions can be readily generalized
to the case of the $SU(2)$ circular solution with two unequal spins
\ci{art}:
X$_1= \cos \psi_0\  e^{ i w_1 \t + i k_1 \s},$ \
 X$_2=\sin \psi_0 \   e^{ i w_2 \t - i k_2 \s},$ \
where  $k_1 J_1 = k_2 J_2, \ J= J_1 + J_2 $
 and  the classical energy has the expansion
 (we assume $k_1,k_2 > 0$)
\be \la{kk}
E=J  + { \l \ov 2 J^2} ( k^2_1 J_1 + k^2_2 J_2)  +
 O({\l^2\ov J^3})=
 J  + { \l \ov 2 J^2}  M^2  +  O({\l^2\ov J^3})  \ ,\ee
 \be
 M^2 \equiv  m^2 \a ( 1- \a) \ , \ \ \ \ \
 m \equiv  k_1 + k_2 \ , \  \ \ \ \ \  \a\equiv {J_2 \ov J} \ .
\ee
The above equal-spin case of $k_1=k_2=k, \ J_1=J_2$
corresponds to
$\a= \ha, \ m= 2 k, \ M= k$.
The generalizations of \rf{z},\rf{jo} and \rf{ha},\rf{po}
are found to be
\be \la{joo}
E_{\rm zero} = E_{\rm reg} + O( { \l^2 \ov J^4})\ ,
\ \ \ \ \ \      E_{\rm reg} =
{ \l  \ov 2 J^2} [ \sum^{\infty}_{n=1}
 ( n^2 - 2 M^2 )]_{\rm reg} =  { \l  \ov 2 J^2} M^2  \ ,
 \ee
 \be \la{poo}
E_{\rm non-zero} = E_{\rm fin} + O( { \l^2 \ov J^4})\ ,
\ \ \ \ \ \ \ \
 E_{\rm fin}= { \l  \ov 2 J^2}
 \sum^{\infty}_{n=1}
 ( n \sqrt{ n^2 - 4 M^2}  -   n^2 + 2 M^2 ) \ . \ee
 This solution is again unstable  for any physical   values
 of the parameters:  the sum is real only for $M^2 < 1/4$
 which is outside the required range.

 Another generalization is
 to the case of the
  3-spin  circular constant-radii solutions of \ci{art}
 belonging to the $SU(3)$ sector.
 In particular, for the case of the
 solution  with $J_1=J_2, \ J_3\not=0$, \ $J= 2 J_1 + J_3 $
 (related by an $SO(6)$ rotation to the  solution of \ci{ft2,ft3}
 but again having manifestly integer-labelled  fermionic
  fluctuation modes)
 one readily finds
 the analogs of \rf{joo} and \rf{poo}
  using the expressions in  \ci{ft3,fpt}.
 This solution is stable  for large
 enough $J_3$, so the 1-loop
 correction to the energy is real.
 The classical energy  is
 \be \la{s3}
 E_0 = J + { \l   \ov 2J} k^2 s^2 + O( { \l^2 \ov J^3})  \ , \ \ \ \ \ \
 s^2 \equiv   1 - { J_3 \ov J} \ .
 \ee
 Expanding the fluctuation frequencies given in \ci{ft3,fpt}
 in ${1\ov \J^2}  = {\l \ov J^2}$
 one  observes again that contributions
 of other superstring modes
 combine to make the sum of the corresponding
 \LL frequencies  finite
 (the latter were
 identified \ci{hl1} directly in the $SU(3)$   \LL model on $CP^2$
 \ci{hl1,st}
  and reproduced also  on the $SU(3)$ spin chain side in
 \ci{frey})\foot{The contributions
 of the two \LL frequencies may be combined as in \rf{hi}
  together  using the identity
 $\sqrt{ a- b} + \sqrt{ a+ b} = \sqrt{ 2 a^2 + 2 \sqrt{ a^2
  - b^2}}$.  This solution is
 stable for $s^2 \leq  1 - (1 - { 1 \ov 2k})^2$.
 To leading order in $1/\J$  the parameter
  $s^2$ is the same as the
 parameter $q = \sin^2 \g_0$ used in \ci{ft3}.
 }
 \be
 \la{su33}
  E^{(1)}_{\rm zero} = E_{\rm reg} =
{ \l  \ov  2J^2} [ \sum^{\infty}_{n=1}
 ( - 2n^2 + 2 k^2 s^2 )]_{\rm reg} =  { \l  \ov 2 J^2} k^2 s^2
   \ ,\ee
 \bea \la{su3}
 E^{(1)}_{\rm non-zero} = E_{\rm fin}&=&{ \l \ov 2 J^2}
 \sum^{\infty}_{n=1}
 \bigg[ n \sqrt{ n^2  + 2 ( 2 -3 s^2) k^2  +
 2k \sqrt{  4n^2(1-  s^2)  -  k^2 s^2 (8 - 9s^2)}} \nn \\
 &+&  n \sqrt{ n^2  + 2 ( 2 -3 s^2) k^2  -
 2k \sqrt{  4n^2(1-  s^2)  -  k^2 s^2 (8 - 9s^2)}} \nn \\
 &-&   2n^2 + 2 k^2 s^2  \bigg] \ .
  \eea
  This reduces back to   \rf{z},\rf{ha} or
  \rf{fi},\rf{po}  for $J_3=0$,
  i.e.~$s=1$\footnote{The above expression can be readily generalized
to the case of the generic  circular solution of \ci{art} with three
unequal spins $J_i$. Computation of the 1-loop  string effective
action near such general solution was recently discussed by H. Fuji
and Y. Satoh (to appear).}.
 We see again that  zero-mode
 contribution  is the same as the $\zeta$-function regularized
 singular part  of the sum of the \LL fluctuation
 mode
 contributions, and that  again  $ E_{\rm reg} $
 is simply  the classical term in \rf{s3} times $1/J$.
 This suggests that it should come out of the non-anomalous
  finite-size correction in the $SU(3)$ spin chain generalization
  of the $SU(2)$ computation in \ci{lz}.\foot{This was indeed
  confirmed while this paper was in preparation in
  \ci{fk}.}

 \bigskip

 Essentially the same conclusions were reached in \ci{ptt}
 in the case of  a  rigid circular  solution
 in the $SL(2)$ sector \ci{art}
 carrying spin $S$ in $AdS_5$  and spin $J$ in $S^5$.
 This solution may be viewed  as a ``naive{}''
 analytic continuation of the  above $(J_1,J_2)$  solution
 in the $SU(2)$ sector.\foot{Direct analytic continuation
 in the spirit  of \ci{bfst}  ($(E,S,J) \to (-J_1,J_2,-E)$,
 etc.)
 leads to a problem of periodic time
 coordinate  and thus should be supplemented by an additional
 redefinition of $\t$ and $\s$. Most of the leading-order
 relations for the energy and the
 fluctuation frequencies are still very
 similar, cf.  \ci{art,ptt}.}
 The non-zero of 3+3  \adss  complex embedding coordinates
 are Y$_0 = \cosh \rho_0 \  e^{i \k \t},$ Y$_1= \sinh\rho_0 \
  e^{i w \t + i  m \s},$ X$_1= e^{i \om \t - i k \s}$.\foot{Here we
  interchange the notation for the $AdS_5$ and $S^5$
   winding numbers
  $m \leftrightarrow k$ compared to \ci{ptt}
  (we choose  $m,k >0$).
 We  also use $\a$
   instead of $u$ in \ci{ptt}
   as a notation for the spin ratio $S\ov J$.}
  The classical energy is
\be \la{sl}
E=J + S   + { \l \ov 2 J}  M^2  +  O({\l^2 \ov J^3})  \ ,  \ \ \ \ \ \ee
\be
M^2 \equiv  m^2  \a ( 1 +  \a)\ , \ \ \ \ \a\equiv { S\ov J} \ ,
\ \ \ \ \    m S = k J  \ .    \ee
The expression for the
1-loop string correction to te energy  $E_1 = E_{\rm zero}
+ E_{\rm non-zero}$ which is again the same as the regularized
sum of the  $SL(2) $ \LL fluctuation mode contributions
are the analogs of
 \rf{z},\rf{jo}, \rf{ha},\rf{po}
 which happen to be simply
  \rf{joo},\rf{poo}  with $M^2 \to - M^2$
   \ci{ptt}\foot{We take into account
a slight correction
to the expression in the original version of \ci{ptt}
(to be done in its revised version)
removing the splitting the sum over $n$  into two  parts.}
\be \la{soo}
E_{\rm zero} = E_{\rm reg} + O( { \l^2 \ov J^4})\ ,
\ \ \ \ \ \      E_{\rm reg} =
{ \l  \ov 2 J^2} [ \sum^{\infty}_{n=1}
 ( n^2 + 2 M^2 )]_{\rm reg} =  - { \l  \ov 2 J^2} M^2  \ ,
 \ee
 \be \la{loo}
E_{\rm non-zero} = E_{\rm fin} + O( { \l^2 \ov J^4})\ ,
\ \ \ \ \ \ \ \
 E_{\rm fin}= { \l  \ov 2 J^2}
 \sum^{\infty}_{n=1}
 ( n \sqrt{ n^2 + 4 M^2}  -   n^2 - 2 M^2 ) \ . \ee
 In contrast to  its $SU(2)$ analog, this $(S,J)$ solution is
 always stable, so that $E_{1}$ is  real and can be
 directly compared to the $SL(2)$  spin chain result
 for the finite-size correction
 (that the classical term in \rf{sl} matches
 the leading Bethe ansatz result was already shown  in \ci{kz}).

 This   is  what we are going to  do in the next section.

\section{Bethe ansatz: finite size corrections}

\subsection{The $SL(2)$ sector}

We shall first  consider  the $SL(2)$ sector which is not plagued by
instabilities and for that reason is conceptually simpler. The Bethe
ansatz for $SL(2)$ is also technically simpler because all Bethe
roots are real. The $SL(2)$ sector consists of operators of the form
$\,{\rm tr}\,D_+^SZ^J$. These operators are dual to strings with the
spin $S$ in $AdS_5$ and the angular momentum $J$ in $S^5$. The
spectrum of anomalous dimensions of the $SL(2)$ operators is
described at one loop by solutions of the Bethe equations \cite{bs}:
\begin{equation}\label{sl2BE}
 \left(\frac{u_k-i/2}{u_k+i/2}\right)^J
 =\prod^S_{j\neq k}\frac{u_k-u_j+i}{u_k-u_j-i}\,,
\end{equation}
where $k,j=1,\ldots ,S$ and $J$ plays the role of the spin chain
length.
 The one-loop anomalous dimension is
\begin{equation}\la{en}
 E-S-J=\frac{\lambda }{8\pi
 ^2J^2}\sum_{k=1}^{S}\frac{1}{u_k^2+1/4}\,.
\end{equation}
The cyclicity of the trace in the SYM operators -- equivalent to the
translational invariance of the wave function -- imposes an
additional constraint:
\begin{equation}\label{mcon}
 \prod_{k=1}^{S}\frac{u_k-i/2}{u_k+i/2}=1.
\end{equation}
It is useful to keep in mind that solutions of the Bethe equations
that do not satisfy this condition still correspond to perfectly
well-defined eigenstates of the underlying spin chain. They carry a
non-zero total momentum and thus have no interpretation in the SYM
theory.

In order to take the large-$J$ limit we rescale
 the Bethe roots as
$u_k=Jx_k$ ($x_k$ remain finite at $J\rightarrow \infty $), take the
logarithm of both sides of the Bethe equation and expand in $1/J$:
\begin{equation}\label{bebe}
 2\pi m-\frac{1}{x_k}=\frac{1}{i}\sum^{S}_{j\neq k}
 \ln\frac{x_k-x_j+i/J}{x_k-x_j-i/J}\,.
\end{equation}
The omitted terms are of order $O(1/J^2)$ and therefore affect
neither the leading order nor the $1/J$ correction. An arbitrary
phase $2\pi m$ arises because of the arbitrariness in choosing the
branch of the logarithm and,  in principle,
 could be different for
different roots. Requiring that the phase is the same for all roots
is a strong restriction and singles out a particular class of states
\cite{kz}. These states are dual to those string  solutions  whose
fluctuation spectrum is discussed in the previous sections. Let us
define the branch of the logarithm in (\ref{bebe}) with the help of
the integral representation:
\begin{equation}\label{bebe1}
 2\pi m-\frac{1}{x_k}=2\sum_{j\neq k}^{}\int_{0}^{1/J}
 d\varepsilon
 \, \frac{x_k-x_j}{(x_k-x_j)^2+\varepsilon ^2}\,.
\end{equation}
This completely fixes the ambiguity in the definition of the  mode
number $m$.

Formally, the logarithm on the right hand side of (\ref{bebe}) can
be also expanded in $1/J$. The expansion is accurate for most of the
Bethe roots, because normally $x_k-x_j\sim x_k\sim O(1)$, but for a
small fraction of nearby roots with $|k-j|\ll J$ the expansion
breaks down since then $x_k-x_j$ is of order $1/J$. The local
contribution produces an {\it anomaly}
 which affects the $O(1/J)$ corrections
and which has been overlooked in the previous analyses. We shall
calculate the anomaly by extending the approach to finite-size
corrections developed in \cite{lz}.

 In order to solve the Bethe equations in the thermodynamic limit we
introduce the resolvent
\begin{equation}
 G(x)=\frac{1}{J}\sum_{k=1}^{S}\frac{1}{x-x_k}\,.
\end{equation}
It has the following asymptotics at infinity:
\begin{equation}
 G(x)=\frac{\alpha }{x}+\ldots \qquad (x\rightarrow \infty ),
\end{equation}
where
\begin{equation}
 \alpha =\frac{S}{J}\,.
\end{equation}
The total energy and the total momentum are Taylor coefficients of
$G(x)$ at zero:
\begin{equation}\label{PandE}
 P=-G(0),\qquad E-S-J=-\frac{\lambda }{8\pi ^2J}\,G'(0).
\end{equation}
The momentum condition (\ref{mcon}) requires $G(0)$  to be an
integer multiple of $2\pi$.

The Bethe equations can be written in the scaling limit entirely in
terms of the resolvent. The derivation proceeds as follows. Let us
multiply both sides of (\ref{bebe1}) by $1/(x-x_k)$ and sum over
$k$. Observing that
\begin{equation}
 \sum_{j\neq k}\frac{1}{x-x_k}\,
 \frac{x_k-x_j}{(x_k-x_j)^2+\varepsilon ^2}
 =\frac{J^2}{2}\,G^2(x)+\frac{J}{2}\,G'(x)
 -\frac{1}{2}\sum_{j\neq k}
 \frac{1}{x-x_k}\,\frac{1}{x-x_j}\,
 \frac{\varepsilon ^2}{(x_k-x_j)^2+\varepsilon^2}\,,
\end{equation}
we find:
\begin{eqnarray}\label{av}
 G^2(x)-\left(2\pi m-\frac{1}{x}\right)G(x)+\frac{G(0)}{x}
 \eq\frac{1}{J}\left[\sum_{j\neq k}
 \frac{1}{x-x_k}\,\,\frac{1}{x-x_j}\,
 \int_{0}^{1/J}d\varepsilon \,
 \frac{\varepsilon ^2}{(x_k-x_j)^2+\varepsilon^2}\,
 \right.
 \nl
\qquad\left.
 \vphantom{ \frac{1}{x-x_k}\,\frac{1}{x-x_k}\,
 \int_{0}^{1/J}d\varepsilon \,
 \frac{\varepsilon ^2}{(x_k-x_j)^2+\varepsilon^2}\,}
 -G'(x)\right].
\end{eqnarray}

Let us now take $J\rightarrow \infty $. Then only
 $x_k-x_j\sim 1/J$ contribute to the sum on the right hand side.
Hence, the sum is dominated by the local distribution of Bethe
roots, which is approximately linear:
\begin{equation}\label{loc}
x_k-x_j\approx \frac{k-j}{J\rho (x_k)}\,,
\end{equation}
where $\rho (x)$ is the macroscopic density:
\begin{equation}
 \rho (x)=\frac{1}{J}\sum_{k=1}^{S}\delta (x-x_k)=
 \frac{1}{2\pi i}\left(G(x+i0)-G(x-i0)\right).
\end{equation}
Thus,
\begin{eqnarray}\label{sum1}
 \sum_{j\neq k}
 \frac{1}{x-x_k}\,\,\frac{1}{x-x_j}\,\,
 \frac{\varepsilon ^2}{(x_k-x_j)^2+\varepsilon^2}
 &\approx&
 \frac{1}{(x-x_k)^2}\sum_{n\neq 0}^{}
 \frac{\varepsilon ^2J^2\rho ^2(x_k)}
 {n^2+\varepsilon ^2J^2\rho ^2(x_k)}
 \nonumber \\
 &=&\frac{\pi \varepsilon J\rho (x_k)
 \coth (\pi \varepsilon J\rho (x_k)) -1}
 {(x-x_k)^2}\ ,
\end{eqnarray}
and finally we get:
\begin{equation}\label{loop}
 G^2(x)-\left(2\pi m-\frac{1}{x}\right)G(x)+\frac{G(0)}{x}
 =\frac{1}{J}\int_{}^{}\frac{dy\,\tilde{\rho }(y)}{(x-y)^2}\,,
\end{equation}
where
\begin{equation}\label{deftilde}
 \tilde{\rho }=\frac{1}{\pi }\int_{0}^{\pi \rho }
 d\xi \,\ \xi \ \coth\xi\ .
\end{equation}

The equation (\ref{loop}) can be solved perturbatively in $1/J$:
\bea
 G(x)=\pi m-\frac{1}{2x}
 &-&\frac{\sqrt{(2\pi mx-1)^2-8\pi m\alpha x}}{2x}\nn \\
 &-&\frac{1}{J}\,\,\frac{x}{\sqrt{(2\pi mx-1)^2-8\pi m\alpha x}}
 \int_{}^{}\frac{dy\,\tilde{\rho }(y)}{(x-y)^2}\,.
\eea
The momentum condition, $G(0)=-2\pi k$, requires $m\alpha=k $ to be
an integer so that $mS=kJ$. This is the same as in string theory,
but we can consider states with any $\alpha $ as far as the spectrum
of the spin chain is concerned. However, the solutions with
irrational $\alpha $ do not correspond to any operators in
$\mathcal{N}=4$ SYM.

Using (\ref{PandE}) we find for the order $\l$ terms in the
energy \rf{en}
\begin{eqnarray}
 E_0&=&\frac{\lambda m^2\alpha (1+\alpha )}{2J}\,,
 \\ \label{Eodin}
 E_1&=&-\frac{\lambda }{8\pi ^2J^2}
 \int_{}^{}\frac{dx\,\tilde{\rho }(x)}{x^2}\,,
\end{eqnarray}
where the effective density $\tilde{\rho} $ is defined in
(\ref{deftilde}). The true density is given by
\begin{equation}
\rho (x)=\frac{\sqrt{8\pi m\alpha x-(2\pi mx-1)^2}}{2\pi x}\,.
\end{equation}
Interchanging the order of integrations in (\ref{Eodin}),
(\ref{deftilde}) and rescaling the integration variable we get:
\begin{equation}\label{E1BA}
 E_1=-\frac{2\lambda M^3}{ J^2}
 \int_{-1}^{1}dx\,x\sqrt{1-x^2}\,
 \coth\left(2\pi M \,x\right) , \ \ \ee
 \be
 M\equiv m\sqrt{\alpha (1+\alpha ) }\ .
\end{equation}
This is our final result.

The integral in (\ref{E1BA}) cannot be expressed in elementary
functions, but we can easily find its asymptotics at
small and
large filling fraction $\a$:
\begin{equation}\label{c}
 E_1=-\frac{\lambda m^2}{2J^2}
 \left[\alpha +\left(1+\frac{\pi ^2m^2}{3}\right)\alpha ^2
 +\ldots \right]\qquad (\alpha \rightarrow 0),
\end{equation}
\begin{equation}\label{a}
 E_1=-\frac{4\lambda m^3\alpha ^{3}}{3 J^2 }+\ldots\qquad
 (\alpha\rightarrow \infty  ).
\end{equation}

To compare to the string theory result let us convert the finite
integral in \rf{E1BA} into a sum using  the identity \foot{This is
equivalent to undoing the summation in (\ref{sum1}). It is
interesting that the mode numbers of the frequencies ($n$'s) in the
Bethe ansatz have the meaning of the distances between the roots
along the contour ($k-j$ in (\ref{loc})). We have no explanation for
this fact.}
 \be
 \pi a\coth ( \pi  a )  = a^2  \sum^\infty_{n=-\infty}
{ 1 \over n^2 + a^2 }
= 1
 +  2 a^2  \sum^\infty_{n=1}  { 1 \over n^2 + a^2 }\ ,
\ee with $a= 2 M x$ in \rf{E1BA}. Then doing the integral over $x$
first we reproduce the string-theory result \rf{soo},\rf{loo} of
\cite{ptt}, i.e.
\be \la{str}
 E_1 = - { \l M^2  \ov 2J^2 }
 +  { \l \ov 2J^2 }
 \sum_{n=1}^\infty  \bigg( n \sqrt{ n^2 + 4 M^2 }
                                     - n^2 - 2 M^2 \bigg)
  +  O (  { \l^2 \ov J^4 })  \ , \ee
where the first term is the contribution of the zero
modes.\footnote{It is possible to do the same calculation in a
different way. First we write the integral over $ x$ as a contour
integral around the cut of the square root and then deform the
contour
 to encircle the poles
 of $\coth (2 \pi M x)$ at $x=\pi n i/2M$.
 This leads to a divergent series
because the
 integrand does not fall sufficiently fast
at infinity and therefore it is necessary to do subtractions before
deforming the contour. Subtracting and adding  $x^2-1/2$ from
$x\sqrt{x^2-1}$  gives the finite sum over the non-zero modes,
 the subtraction can be evaluated by shrinking
the contour to zero where $\coth$ has a pole and this produces a
zero-mode contribution.}


\subsection{The $SU(2)$ sector and instabilities}

The Bethe equations for the $SU(2)$ operators
\ ${\mathop{\mathrm{tr}}\nolimits}(Z^{J_1}W^{J_2}+{\rm
permutations})$ \cite{mz} differ from their $SL(2)$ counterpart
(\ref{sl2BE}) by reversing the signs on the left hand side:
\begin{equation}\label{besu2}
 \left(\frac{u_k+i/2}{u_k-i/2}\right)^J
 =\prod^{J_2}_{j\neq k}\frac{u_k-u_j+i}{u_k-u_j-i}\,,
\end{equation}
where $J=J_1+J_2$. All the previous calculations for  the $SL(2)$
case can be literally repeated in this case. The only subtle point
is eq.~(\ref{loc}), because the roots are now complex. But the
density is also complex -- it is the product $dx\,\rho (x)$ that
must be real and positive -- so (\ref{loc}) holds even if the roots
lie on the complex plane. In fact,
 we need
not do separate calculations for the
$SU(2)$ case since the one-loop
anomalous dimensions in the the $SU(2)$ and $SL(2)$ sectors are
related by the
analytic continuation in the filling fraction. This fact
was established for the thermodynamic limit in \cite{bfst} and holds
true for the leading $1/J$ correction as well. If we define
\begin{equation}
 \alpha =\frac{J_2}{J}=\frac{J_2}{J_1+J_2}\,,
\end{equation}
then the finite-size correction in  the $SU(2)$
case  can be obtained from
(\ref{E1BA}) by the substitution $\alpha \rightarrow -\alpha $:
\begin{equation}\label{E1BA'}
 E_1=\frac{2\lambda M^3}{ J^2}
 \int_{-1}^{1}dx\,x\sqrt{1-x^2}\,
 \cot\left(2\pi M \,x\right) , \ \ \ee
 \be
 M\equiv m\sqrt{\alpha (1-\alpha ) }\ .
\end{equation}
This formula makes sense only if $M$ is sufficiently small,
and then all
the poles of the integrand lie outside of
 the region of integration.
If we analytically continue $E_1$ past
\begin{equation}\label{alc}
\alpha _c=\frac{m-\sqrt{m^2-1}}{2m}\ ,
\end{equation}
the poles hit the contour of integration and $E_1$ acquires an
imaginary part from the residue. The poles are associated with the
frequencies of fluctuations around the corresponding solution of the
Landau-Lifshitz equation, and the imaginary frequency signals that
the solution becomes unstable. Note that the momentum condition
requires integrality of $m\alpha =k$ and implies that $M\geq 1$, so
the string states dual to the SYM operators are always unstable, in accord
with the analysis of \cite{ft2,art}.

How do we see this instability in the Bethe equations? It is easy to
understand what goes wrong. The density of Bethe roots grows with
$\alpha $ and at $\alpha =\alpha _c$ some roots appear to be
separated by $i/J$. This causes problems in the calculation of the
anomaly -- the sum in (\ref{sum1}) becomes ill-defined since the
denominator of the $n=1$ term turns to zero. In fact, the solutions
with $\alpha
>\alpha _c$ violate the basic assumption used in deriving the macroscopic
equations for the distribution of Bethe roots, namely the assumption
that all logarithms $\ln (x_k-x_j+i/J)/(x_k-x_j-i/J)$ belong to the
same branch. At the leading order only well-separated roots are
important. The argument of the $\ln$  is close to one for them and
it is easy to forget about this assumption. Indeed,  the derivation of
scaling solutions of the Bethe equations, which are dual to the
semiclassical string states \cite{bmsz,bfst,kmmz}, starts with
rewriting the microscopic equations in the logarithmic form:
\begin{equation}\label{belog}
 \sum_{j\neq k}^{}
 \ln\frac{x_k-x_j+i/J}{x_k-x_j-i/J}
  =2\pi im_k+J\ln\frac{x_k+i/2J}{x_k-i/2J}\,,
\end{equation}
where $x_k=u_k/J$. Then these equations  are expanded in $1/J$
 and rewritten as an integral equation for the density of roots:
\begin{equation}\label{macros}
 2\pint \frac{dy\,\rho (y)}{x-y}
 =2\pi m_I+\frac{1}{x}\,,\qquad x\in C_I.
\end{equation}
The density is supported on a set of contours $C_I$ in the complex
plane. The integral equation can be now solved and its general
solution can be expressed in terms of hyperelliptic integrals
\cite{kmmz}. This derivation tacitly assumes that the logarithms in
(\ref{belog}) are single-valued and that all roots with the same
phase lie on the same contour $C_I$. While this assumption is
certainly justified for well-separated roots, it can break down if
$x_k-x_j=O(1/J)$. For such $x_k$ and $x_j$ the local approximation
(\ref{loc}) is accurate and the logarithm in (\ref{belog}) takes the
form
\begin{equation}
 \ln\frac{n+i\rho (x)}{n-i\rho (x)}\equiv F_n(x),
\end{equation}
where $n=k-j$. We should require that $F_n(x)$ is single-valued
along each $C_I$. Otherwise the phase $2\pi m_I$ will jump by an
integer multiple of $2\pi $ somewhere in the middle of the contour.
This extra condition does not follow from the macroscopic equations
(\ref{macros}) themselves and should be imposed by hand.
 In effect, only those solutions of the classical
Bethe equation correspond to microscopic Bethe states which satisfy
the following

{\bf Stability condition:} {\it The density of roots must satisfy}
\begin{equation}\label{stab}
 \Delta_{C_I} \arg\frac{n+i\rho (x)}{n-i\rho (x)}
 =0
\end{equation}
{ \it for any integer $n$ and for all contours $C_I$. The density is
zero at the ends of all cuts, so $(n+i\rho(x) )/(n-i\rho(x) )$
traverses a closed curve in the complex plane which begins and ends
in $1$. The solution is stable if this curve does not encircle the
origin.}

For many interesting solutions there is only one point ($x=x_*$) at
which $dx$, and thus $\rho (x_*)$, is pure imaginary and
consequently the curve $(n+i\rho(x) )/(n-i\rho(x) )$ crosses the
real axis only once, at $x=x_*$. In that case the solution is stable
iff the crossing point lies to the right of the origin, or if
\begin{equation}\label{stab1}
 |\rho (x_*)|<1.
\end{equation}
This condition has a simple meaning: when the difference between
adjacent roots is pure imaginary, it should be smaller than $i/J$.
This makes the logarithm in (\ref{bebe}) single-valued.

We conjecture that solutions of the classical Bethe equation that
violate the stability condition correspond to unstable solutions of
the Landau-Lifshitz equation. Consider, for example,
the rational (single-cut)
solution:
\begin{equation}\label{rhora}
 \rho (x)=\frac{i\,\sqrt{8\pi m\alpha x+(2\pi m x-1)^2}}{2\pi x}\,.
\end{equation}
The density is supported on a single contour which crosses the real
axis at
\begin{equation}
x_*=\frac{1}{2\pi m(1-2\alpha )}\,,
\end{equation}
and
\begin{equation}
\rho (x_*)=2mi\alpha \sqrt{1-\alpha }\,.
\end{equation}
The stability condition (\ref{stab1}) demands
\begin{equation}
 \alpha <\alpha _c,
\end{equation}
where $\alpha _c$ is defined in (\ref{alc}), which is precisely the
condition for linear stability of the corresponding classical
solution.

The stability condition is very similar to the consistency
condition for the Douglas-Kazakov solution of large-$N$ QCD on a
two-dimensional sphere \cite{Douglas:1993ii}, where the density is
also bounded from above. When the bound is saturated, the solution
undergoes a phase transition and develops a patch with flat
distribution. Similar phase transition will occur here. If  the
stability bound gets violated on a contour $C_I$, for example,
 if the
filling fraction of the rational solution (\ref{rhora}) exceeds the
critical value $\alpha _c$, roots on different parts of $C_I$ will
have different mode numbers $m_K$ and therefore the contour will
break in two pieces on which the mode numbers are constant. For
instance, the rational solution (\ref{rhora}) will become two-cut.
The two cuts will be connected by a condensate in which Bethe roots
are exactly equidistant: $x_{k+1}-x_k=i/J$. The solutions with such
condensates are discussed in more detail in \cite{bmsz,bfst,kmmz}.

Finally, let us write down the $1/J$ correction to the classical
Bethe equation (\ref{macros}):
\begin{equation}\label{macros1}
 2\pint \frac{dy\,\rho (y)}{x-y}
 =2\pi m_I+\frac{1}{x}-\frac{1}{J}\,\pi \rho '(x)\coth\pi \rho (x),
 \qquad x\in C_I.
\end{equation}
One derivation is given in Appendix~\ref{appsadlkfj}. The other can proceed by
deriving an analog of eq.~(\ref{loop}) for arbitrary mode numbers
and then taking its discontinuity across the cuts $C_I$. We only
know how to solve for the $1/J$ corrections in the simplest case of
the rational solutions, but it is conceivable that
 the finite-size corrections can be computed in the same generality in
 which the
leading-order solution is known \cite{kmmz}.

 Perhaps (\ref{macros1}) can be solved by
iterations.\footnote{We would like to
thank I.~Kostov for the discussion of this point.}
 The leading order is known.
 Solving for the next iteration
  reduces to inverting the Hilbert kernel,
 which can be done by fairly standard techniques
\cite{gakhov}.

\section{Concluding remarks}

It has been observed by direct computations that the energies of the
semiclassical strings agree
 with the anomalous dimensions at two gauge-theory
loops and start
 to disagree at three loops. This is true for
classical macroscopic strings, as well  for short strings in the
near-BMN limit. We believe that quantum string/finite-size
corrections should be no exception and that they should agree at two
loops. The two-loop calculation on the gauge-theory side is
certainly possible, since the two-loop Bethe ansatz for the $SL(2)$
sector is known \cite{s}. In fact, the two-loop corrections do not
affect the anomaly and only change the macroscopic part of the
classical Bethe equation. The expansion of the string one-loop
quantum correction
 to the ``gauge{}''
two-loop order
 $O(\lambda^2/J^4)$
meets certain technical difficulties. The sum of the
frequencies expanded in $\lambda /J^2$ diverges at $O(\lambda
^2/J^4)$ and it is necessary to first resum the series and
 then
expand. One can of course resort to numerical evaluation of the sum.

Another obvious extension of our results is the calculation of the
$1/J$ corrections from the quantum string Bethe ansatz \cite{afs,s},
which was conjectured to describe the string spectrum at strong
coupling but beyond the semiclassical limit. Comparison of the $1/J$
corrections computed from the Bethe ansatz with the explicit string
calculation would be a strong test of the conjecture of
\cite{afs,s}.

We also believe that our approach can be generalized to the $SU(3)$
sector. While this paper was in preparation, there appeared an
interesting work \ci{fk} that generalized
 the computation of non-anomalous part of finite-size
 correction in \ci{lz} from $SU(2)$ to $SU(3)$ case.
 As we have explained  above,
 this non-anomalous ``$1\to 1 + { 1 /J}$'' correction
 to the classical energy
 corresponds  to the zero-mode string contribution and
 should again be supplemented by the anomaly,
 after which we expect the  spin-chain result to  agree
 with the string-theory prediction, i.e. the sum of \rf{su33}
 and  \rf{su3}.

Potentially, $1/J$ corrections can be also computed for many other
known solutions for classical strings in $AdS_5\times S^5$,
e.g.,
 for
folded $SU(2)$ string solution
 of \ci{ft4} dual to the
 two-cut solution of  the Bethe ansatz \ci{bmsz}.

An interesting open problem is to show that in general the
finite-size correction is always given by a sum over energies
(obtained by removing one root from a cut \ci{bmsz}) of the nearby
fluctuation modes in a given spin-chain sector, i.e. by the
(regularized) sum of the energies of the Landau-Lifshitz modes.

\section*{Acknowledgments }

We are grateful to
S.~Frolov, I.~Kostov, A.~Mikhailov,  J.~Minahan,
I.~Park and  A.~Tirziu for discussions
and especially V.~Kazakov for sharings his insights
into anomalies within the Bethe equations.
We  thank   R.~Hern\'andez and E.~L\'opez for  discussions
and letting us know about their forthcoming work
\cite{asdlaslkjghsdfh}
on finite-size corrections for the $SU(2)$ spin chain
 with  similar conclusions  obtained by a
 somewhat different method.
 We also  thank G.Korchemsky for drawing our
  attention to ref.\ci{kor} where a semiclassical large  spin
  expansion of $SL(2)_{s=0}$  Bethe equations  was discussed
(for fixed  spin chain length)  from a different point of view.

 The work of N.B.~is supported in part by the U.S.~National Science
Foundation Grant No.~PHY02-43680. Any opinions,
findings and conclusions or recommendations expressed in this
material are those of the authors and do not necessarily reflect the
views of the National Science Foundation.
The  work of A.A.T.  was supported  by the DOE
grant DE-FG02-91ER40690  and also by
 the INTAS contract 03-51-6346
and the RS Wolfson award.
The work of K.Z. was supported in
part by
the Swedish Research Council under contracts 621-2002-3920 and
621-2004-3178, and by the G\"oran Gustafsson Foundation.

\appendix

\section{Anomaly}
\label{appsadlkfj}

In this appendix we demonstrate the effect of the anomaly
on the Bethe equations for Bethe strings in the thermodynamic limit.
In the leading order in $1/J$ the anomaly the anomaly is
absent \cite{asadfasdflsdlfaldsfkjh}, but it does
contribute to subleading orders:
A Bethe string is a collection of Bethe roots $u_n$
distributed on a curve $C$ in the complex plane.
Let us focus on a particular point of this curve,
for convenience we shall assign the index $n=0$ to this point.
Then for large $J$ we can expand the positions of Bethe roots
close to $u_0$ as follows
\begin{equation}
u_n=aJ+b n
+\frac{1}{2}c n^2/J
+\ldots
\end{equation}
where $a,b,c,\ldots$ depend on $J$ but are finite for $J\to\infty$.
The scattering phase for $u_0$ is
\begin{equation}
\phi=
\frac{1}{i}\sum_n\log\frac{u_0-u_n+i}{u_0-u_n-i}
\end{equation}
Substituting the above $u_n$  into $\phi$
 and separating  into small and large $n$ we get
\begin{equation}
\phi=
\frac{1}{i}\sum_{|n|<N}\log \frac{-b n-c n^2/2J+i}{-bn-cn^2/2J-i}
+\sum_{|n|>N}\frac{2}{u_0-u_n}
+\ldots
\end{equation}
Then let us
 sum up terms with positive and
 negative $n$ in the first term
and expand
\begin{equation}
\phi=
\sum^N_{n=1}
\left(
\frac{2c/J}{b^2}
-
\frac{2c/J}{b^2(1+b^2n^2)}
\right)
+\sum_{|n|>N}\frac{2}{u_0-u_n}
+\ldots
\end{equation}
The first term can be absorbed into the last one because
${2\over u_0-u_n}=2c/Jb^2$ for small $n$.
The sum in the second term converges linearly,
thus we can send $N\to\infty$ at this order,
\begin{equation}
\phi=
\sum_{n}\frac{2}{u_0-u_n}
+\frac{c}{b^2J}(1-\frac{\pi}{b} \coth(\frac{\pi}{b}))
+\ldots
\end{equation}
The second term is the anomaly.
The hyperbolic cotangent represents
the effect of the poles of the nearby Bethe roots.
It can be computed using the leading order
density $\rho(x)=dn/dx$ with $xJ=u$.
Then $b=1/\rho(x_0)$ and $c=-\rho'(x_0)/(\rho(x_0))^3$
and thus the anomaly contribution is
\begin{equation}
\delta \phi=
\frac{1}{J}\,\frac{\rho'(x_0)}{\rho(x_0)}
(\pi\rho(x_0) \coth(\pi \rho(x_0))-1)
\end{equation}
We see that it is a purely local term.
Finally, we can approximate the sum
by an integral
\begin{equation}
\phi=
\pint \frac{2\,dx\,\rho(x)}{x_0-x}
+\frac{1}{J}\,\pi\rho'(x_0)
\coth(\pi \rho(x_0))
+\ldots
\end{equation}
The term $\rho'(x_0)/\rho(x_0)$ was absorbed by turning the sum into
an integral.


\end{document}